# Formation of anions and cations via a binary-encounter process in $OH^+$ + Ar collisions: the role of dissociative excitation and statistical aspects

E. Lattouf,[1] Z. Juhász,[2,*] J.-Y. Chesnel,[1,*] S. T. S. Kovács,[2] E. Bene,[2] P. Herczku,[2] B. A. Huber,[1] A. Méry,[1] J.-C. Poully,[1] J. Rangama,[1] and B. Sulik[2]

[1]*Centre de Recherche sur les Ions, les Matériaux et la Photonique (CIMAP), Unité mixte CEA-CNRS-Ensicaen-Université de Caen Basse-Normandie, UMR 6252, 6 Boulevard Maréchal Juin, F-14050 Caen cedex 04, France*

[2]*Institute for Nuclear Research, Hungarian Academy of Science (MTA Atomki), H-4001 Debrecen, P.O. Box 51, Hungary*

*Corresponding authors: zjuhasz@atomki.hu, jean-yves.chesnel@ensicaen.fr

Molecular fragmentation leading to the formation of negatively and positively charged hydrogen ions in 7-keV $OH^+$ + Ar collisions is investigated experimentally. The most striking finding is that negative and positive hydrogen ions are emitted with very similar angular dependences. Also, the kinetic energy distribution of the $H^+$ fragment shows strong similarities with that of the ejected $H^-$ ion. The kinematics of the emitted H core is found to be essentially driven by its scattering on the atomic target. However, in addition to this binary-encounter process, dissociative electronic excitation of the molecular projectile has to be invoked to explain the observed fragmentation patterns. Though the electron capture process is complex, it is shown that the relative population of the different final charge states of the outgoing fragments can be described by simple statistical laws.

**Introduction**

In collisions involving molecular species, highly reactive cations, anions and neutral radicals may be formed. These species have a significant role on the chemistry of astrophysical and biological media. This is particularly true for negative ions which have been a subject of prime interest over the past decades [1-6]. Anions play a major role in many areas of physics and chemistry involving weakly ionized gases and low energy plasmas [1-4,7]. Even in small concentrations, anions influence appreciably the properties of their environment [1-4,8,9]. However, not all the mechanisms of their creation are known.

Hydrogen anions can be formed in collisions between cations and neutral atoms or molecules. Several studies have shown the formation of $H^-$ ions in soft collisions involving negligible momentum transfer between the collision partners [10-16]. However, in a recent study [17], we observed that $H^-$ ions can also be created in hard collisions involving energetic encounters between atomic cores. In $OH^+$ + Ar collisions, the observed $H^-$ ions were formed by a combined process, in which the incoming proton was scattered by the Ar target to large angles and then captured two electrons when leaving the collision complex. In the whole investigated angular range (30°–150°), the kinematics of the $H^-$ fragment could be well described by a simple two-body scattering calculation for the binary collision of the H atom on the Ar target. Similarly, the measured cross section was found to be proportional to the calculated two-body scattering cross section. Hence, it was concluded that the fraction of the scattered H centers which become negatively charged is independent of the scattering angle, and thus, barely depends on the impact parameter and on the momentum transfer. Accordingly, the distribution of the final charge states of the fragments was suspected to be akin to a statistical distribution [17]. Upon this, the question arises whether a similar angular independence applies for the fraction of positively charged fragments, too. One of the main goals of the present work is to address this question.

Another issue came out when we compared our data [17] to the results of Alarcón *et al*. [14] for $H^-$ production in $H_2^+$ + Ar collisions with nearly the same velocity as in our case. Their results were limited to small angles ($< 4°$) so that there was no overlap between the two investigated angular ranges. However, as seen in Fig. 2(b) of Ref. [17], our scaled cross-section curve for H scattering on Ar nicely connects the two sets of results and matches the cross sections measured at small angles (0.3°–3°). It was surprising that the two-body picture could satisfactorily describe such a wide class of collisions, including collisions leading to small scattering angles (soft collisions at large impact parameter) in which the energy transfer is too small to create an $H^-$ ion from the ground state of the $H_2^+$ radical. Therefore, the second goal of the present work was to extend our measurements to smaller scattering angles in order to see whether the measured cross sections follow the calculated ones for our collision system ($OH^+$ + Ar) as well. Measurements at smaller angles are also of high importance because the scattering cross section increases with decreasing angle, so that the majority of the projectile fragments are emitted at small angles. Also, at angles of a few degrees, the comparison between the experimental data and predictions for elastic two-body and three-body scatterings is expected to provide further insight into the fragmentation process itself.

**Experimental method**

The experiments were performed at the ARIBE facility of the Grand Accélérateur National d'Ions Lourds (GANIL) in Caen, France. The same experimental method has been described in a previous work [17]. It consists of a crossed-beam type experiment in the gas phase. The ion beam was delivered by an electron-cyclotron-resonance ion source (ECRIS). In the collision chamber, it interacted with an effusive gas jet before ending up in a Faraday cup. The latter was used for continuous measurement of the ion beam current in order to normalize the cross sections. The fragments emerging from the investigated collisions were selected according to their kinetic energy per charge unit by means of a 45° parallel plate electrostatic analyzer, with an energy resolution of 5%. This spectrometer was fixed on a rotatable ring allowing the selection of the angle of detection with respect to the ion beam direction, with an angular acceptance of 2°. The particles transmitted by the spectrometer were postaccelerated to ~ 1 keV and detected by a channel electron multiplier (channeltron). It is commonly admitted that the channeltron efficiency is about 80% and 90% for 1-keV $H^+$ and $H^-$ ions, respectively [18]. The global counting efficiency depends also on the transmission of the spectrometer (altered by meshes and edge effects) and of the acquisition system. For our setup the global counting efficiency is 25%, determined in earlier experiments by means of reference cross sections for electron emission. Its relative uncertainty is estimated to be about 50%.

At observation angles smaller than 30°, the beam current cannot be measured by the Faraday cup because the spectrometer is in the way of the beam. Instead, for normalization, we used the measured current on the slits located just before the entrance of the collision chamber. We checked that the currents measured on the Faraday cup and on the entrance slits are proportional to each other. The proportionality factor was determined by comparing the count rate for proton emission at 30° normalized to the current on the Faraday cup with the same entity normalized to the entrance slit current. More precisely, the comparison was made according to the pro-tons produced in binary quasielastic collisions, which resulted in a sharp peak centered at an energy close to 412 eV. This indirect measurement of the beam current led to an estimated current uncertainty of 20%.

In order to obtain the correct angular dependence of the cross section, one has to take into account the effective target length seen by the spectrometer. This length is proportional to $1/\sin(\theta)$, where $\theta$ is the observation angle with respect to the beam direction [19]. With a uniformly distributed target gas, the count rate at the angle $\theta$ has to be multiplied by $\sin(\theta)$ to correct for this geometrical feature. The nozzle used to inject the target gas was 5 mm above the collision center, ensuring a high target density. But, within these conditions, the target gas was not uniformly distributed and significant deviation from the $\sin(\theta)$ dependence is expected, especially below 30°. Therefore, an additional measurement was required: for proton signals we moved the nozzle upwards until the count rate became independent of the nozzle's position, asymptotically reaching a minimum. This means that the target gas became uniformly distributed in the interaction region. This happened when the distance between the nozzle and the collision center was about 50 mm (*up* position). At this nozzle position, we measured the proton signal at a few angles in the range from 3° to 30°. Then, knowing the spectrometer geometry and the pressure in the chamber, we were able to deduce absolute cross sections. The obtained results were compared to the proton signal when the nozzle was at 5 mm above the nominal beam axis (*down* position) at the corresponding angles. From the measured *up/down* yield ratios a simple analytical formula has been determined by fitting. For the yields measured in the *down* position of the nozzle, this formula replaces the above $f_{corr}^{up}(\theta) = \sin(\theta)$ correction. It reads $f_{corr}^{down}(\theta) = f_{90}[c + (1-c)\sin(\theta)]$, with the fit parameters $f_{90} = 0.048$ (the *up/down* yield ratio at 90°) and $c = 0.48$. Here $c$ represents a cylindrical jet component, while $(1-c)$ refers to a homogeneous component of the target gas. Though approximate, this formula well reproduces the measured *up/down* ratios. In all cases, when the data were collected in the down position, e.g., for negative ions, the factor $f_{corr}^{down}(\theta)$ was used in determining the cross sections. The estimated statistical uncertainty for this normalization factor is 20%.

Hence, by taking into account the different sources of errors, the uncertainty of the present calibration in absolute scale is about 60%, which is less than it was in our previous work [17], in which the calibration relied on earlier measurements. The present calibration leads to cross sections for H$^-$ production 1.8 times lower than those in Ref. [17]. This factor is just within the estimated uncertainty limits for our earlier results [17], but it is not negligible. Thus, a correction of the previous results is also included in this work.

**Results and discussion**

The present study focuses on ionic fragments emitted at small observation angles. In order to compare the measurements in our current study with the previous results, some data have also been recorded at larger angles, up to 90°. The energy spectra of negative and positive H ions formed in 412 eV/amu OH$^+$ + Ar collisions are shown in Figs. 1 and 2. These spectra were recorded at emission energies ranging from 150 to 800 eV. This range is wide enough to contain the peak structure due to the removal of hydrogen ions since these ions are expected to have a mean kinetic energy of ~ 412 eV in the laboratory frame (at small forward angles).

As is visible in Fig. 1 for negative ions and in Fig. 2 for positive ions, the spectra cannot be described as a single Gaussian function with a background. Moreover, attempts to correctly fit a single non-Gaussian shape peak function to the main peak for all angles were unsuccessful. Instead, a good fit to the measured data was obtained when including two Gaussian functions in the fitting procedure: a narrow component [~60 eV full width at half maximum (FWHM)] and a wide one (~140 eV FWHM) with slightly different centroids. We used a simple power law function (with adjustable parameters) to fit the background due to electrons. This function appears as a straight line on log-log scale plots, as seen in Figs. 1 and 2.

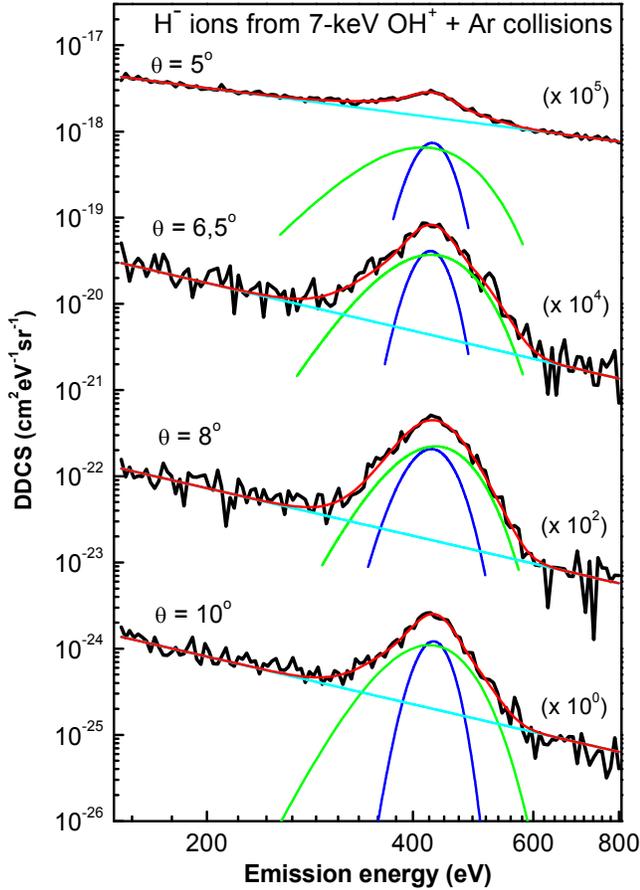

**Figure 1.** (Color online) Black curves: double differential cross section (DDCS) for $H^-$ emission from 7-keV $OH^+$ + Ar collisions at the different observation angles $\theta$ indicated in the figure. The $H^-$ ions produced by binary collisions exhibit a merged double peak structure around 412 eV. The two Gaussian functions used in the fitting procedure are presented (dark blue curve: narrow peak component, green curve: wide peak component, light blue curve: power law background function, red curve: sum). For graphical reasons, each spectrum is multiplied by the factor indicated on the right hand side.

It should be noted that the composition of the peak depends on the observation angle: while negligible at angles larger than 30°, the relative contribution of the wide peak component increases when decreasing the angle down to a few degrees. It is tempting to invoke two different processes to interpret the double structure of the peak. However, our model calculations (introduced later) show that such a structure can be formed merely by the disturbance caused by the third body (oxygen core) on the binary-collision driven $H^{(-/+)}$ emission.

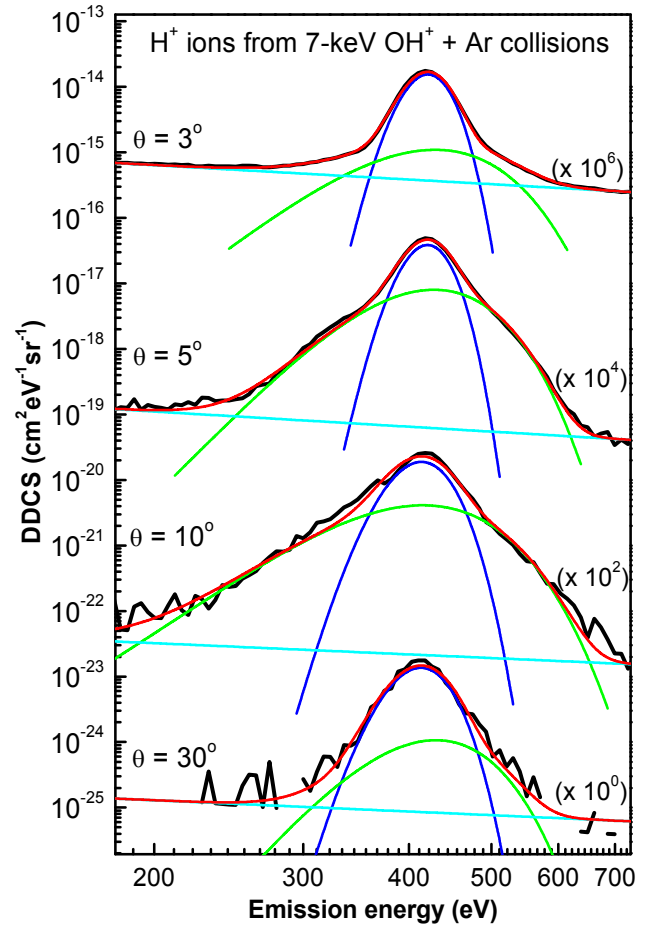

**Figure 2.** (Color online) Same as Fig. 1 but for $H^+$ ions.

The mean energy of the $H^{(-/+)}$ ion emission as a function of the observation angle is an informative quantity about the collision kinematics. It was determined from the fitting parameters as the weighted average of the cen-troids of the two Gaussian components (when double Gaussians were used). As can be seen in Fig. 3 for both the negative and positive ions, the measured mean energies follow the elastic two-body scattering calculation (H colliding on Ar) above 30°.

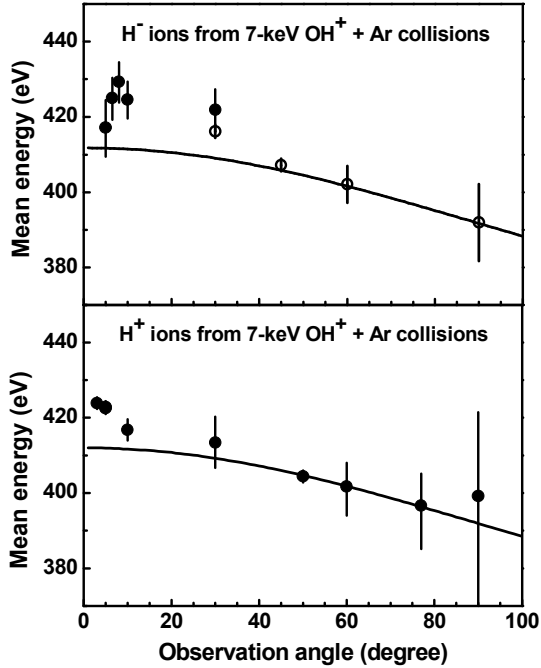

**Figure 3** Average kinetic energy of $H^-$ (top) and $H^+$ (bottom) ions resulting from binary encounter collisions between the H atom of the projectile and the Ar target atom. The energy was determined by the average of the peak positions of the narrow and wide peak components shown in Figs. 1 and 2, weighted by the corresponding cross sections. Filled circles: present results. Open circles: earlier results from [17]. Solid line: predicted final energy of the H ions assuming a pure two-body elastic scattering of a 412-eV H atom on Ar [17].

Below 30°, however, significant deviations are found (Fig. 3). The measured data exceed by 10-20 eV the kinetic energy calculated for a purely elastic collision with a $Q$ value equal to 0 ($Q$ is the amount of energy released by the collision). Since electron transfer processes from the target to the projectile fragments are predominantly endothermic in the present collision ($Q < 0$), these results are rather unexpected. They can be explained by assuming that a small part of the kinetic energy of the incident $OH^+$ ion is converted into electronic excitation to form dissociative states. This would result in a kinetic energy release (KER), which separates the O and H atoms. Since the excitation energy is small compared to the initial kinetic energy of the projectile, the velocity of the $OH^+$ ion in the laboratory frame remains almost unchanged after its excitation. Moreover, because of the relatively large mass of the O atom, the H core takes most of the kinetic energy released during the dissociation of $OH^+$. As a net result, taking into account the random orientation of the $OH^+$, the kinetic energy of the H core is higher on average than the expected value given by the classical two-body elastic scattering calculations.

It has to be mentioned that the emission of H fragments at angles of a few degrees, which corresponds to large impact parameter collisions, cannot be described in terms of a pure elastic scattering of the H center on the Ar atom. For instance, at a scattering angle of 5°, the kinetic energy transferred to the H center in a pure two-body elastic collision with the Ar target is only 3.1 eV in the projectile frame. Since this energy is smaller than the dissociation energy of the $OH^+$ ion in its ground state (5 eV) [20], it is not sufficient to create the observed H ions. Therefore, the emission of H fragments at small angles would not be possible without the dissociative electronic excitation of the $OH^+$ ion.

The singly differential cross sections for $H^-$ and proton emission as a function of the observation angle were determined from the area of the fitting Gaussian functions (from the sum of the two components). As shown in Fig. 4, the angular dependence of both $H^-$ and $H^+$ cross sections is highly similar to that of theoretical cross sections for elastic scattering of H on Ar. (Details on the scattering cross-section calculation are given in our previous paper [17]). When multiplied by an appropriate factor, the theoretical curves match fairly well the experimental data in the entire angular range (Fig. 4). The largest deviations are found for protons, but they are less than 50% of the scaled theoretical values. They may partially stem from the uncertainties of the calibration process.

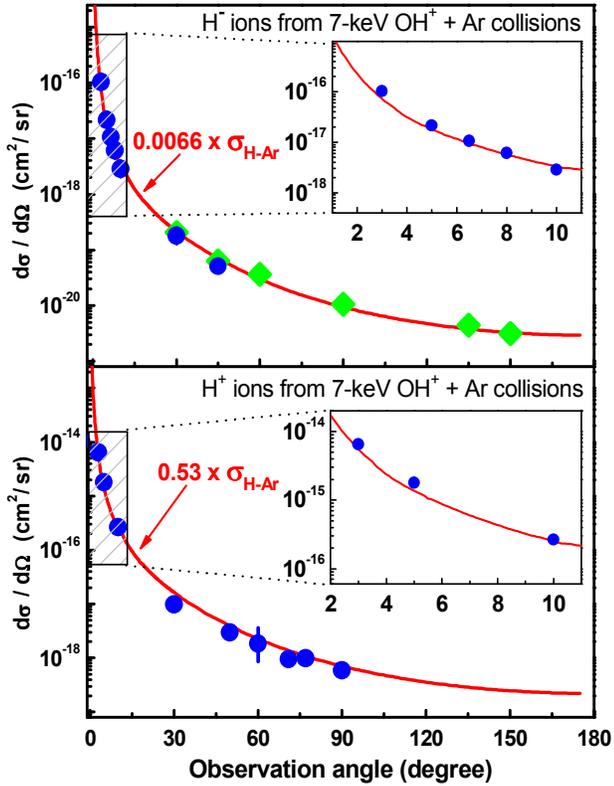

**Figure 4** (Color online) Single differential cross sections (SDCS) for H⁻ (top) and H⁺ (bottom) emission as a function of the observation angle (blue circles). Only *relative* error bars (due to statistical uncertainties and due to the sin($\theta$) correction) are shown. These relative error bars (typically 25%) are smaller than the symbol size, except at 60° for protons, where the large uncertainty stems from the fitting of accidentally overlapping peak structures. Green diamonds: earlier results from [17] with the present normalization procedure. Red curves: calculated cross section for two-body elastic scattering of 412-eV H on Ar, multiplied by factors representing the fraction of the different charge state components.

The physical meaning of each multiplication factor reported in Fig. 4 is the probability to populate the corresponding charge state. Namely, it is found that about $(0.7 \pm 0.4)$% of the scattered H atoms become negatively charged and about $(53 \pm 32)$% of them become positively charged. One can infer that the remaining fraction of about 46% corresponds to neutral H fragments.

The ratio of H⁻ to H⁺ cross sections was found to be ~0.012 in average. Fluctuations around this mean value do not exceed the statistical error bars (60% at maximum). Thus, we may state that the H⁻ to H⁺ ratio is constant in the investigated angular range. A similar behavior is expected for the H⁰/H⁺ production ratio. Earlier, this ratio was measured by Martínez and Yousif for 1-5 keV $H_2^+$ + Ar collisions at observation angles ranging from 1.6° to 7° [21]. They found it to be somewhat higher than unity for 1 keV impact energy, in the angular range above 3° [21]. In the present measurement, in the same angular range, we found indirectly the H⁰/H⁺ ratio to be similar, namely $0.9 \pm 0.7$.

We emphasize that the relative populations of the different charge states do not depend significantly on the scattering angle. Hence, one may expect the charge-state distribution to follow a simple statistical law. One should keep in mind here that the neutral yield has not been measured directly. It is determined from one calculated and two measured cross sections, the latter's with rather large uncertainties. Though the statistical character of the final charge-state distribution of the hydrogen fragments is clear, the uncertainty of the value of their neutral fraction leaves space for different statistical models.

To interpret our findings in statistical terms, we consider the population of the different charge states of the H fragments as a two-step process of electron capture by the atomic core H⁺. This two-step model can be associated, *e.g.*, with an over-barrier picture [22] applied along the outgoing path of the scattered proton, when it leaves the collision complex. Let $p$ denote the probability of capturing one electron from the collision complex, and let $q$ denote the probability of capturing a second electron if one electron has already been captured. The expressions of the probabilities for the possible out-comes of the two-step process are given in Table 1.

Table 1 Probabilities of the two-step capture processes, final charge states and their measured fraction.

| Captured electrons 1st step | 2nd step | Probability | Final charge state | Measured fraction (%) |
|---|---|---|---|---|
| 0 | 0 | $(1-p)^2$ | $H^+$ | 53 ± 32 |
| 1 | 0 | $p(1-q)$ | $H^0$ | - |
| 0 | 1 | $(1-p)p$ | | |
| 1 | 1 | $pq$ | $H^-$ | 0.66 ± 0.40 |

From the knowledge of the $H^-$ to $H^+$ cross-section ratio and of the fractions of the different charge states of the hydrogen fragments, the probability $p$ was estimated to be between 0.08 and 0.54 when taking into account the experimental uncertainties. Similarly, the probability $q$ was found to be between 0.012 and 0.08. These values suggest that the capture of the second electron is less probable than the capture of the first one. Accordingly, formation of a very specific system such as a strongly electron-correlated $H^-$ anion in the ground state may be less likely than formation of neutral H fragments either in the ground state or in excited states. We note, however, that the measured data also allow the $p=q$ approach as an extreme within the experimental uncertainties. This leads to the somewhat unlikely model of two independent capture events, represented by a binomial distribution.

We can conclude that the experimental data do not conflict with the interpretation in terms of a two-step electron capture process. However, it is more likely that the actual process involves many steps at different level crossings, with sets of capture and recapture events. The complexity of the system far exceeds what can be currently investigated. Therefore, no attempt is made for a fully detailed analysis in this work.

**Simulations**

Although rigorous treatment of the capture process is not achievable presently, we performed some simplistic numerical simulations for the trajectories of the different fragments in order to interpret the data presented here. Since these simulations are based on rather crude approximations, only qualitative agreement with the experiments can be expected. We assumed a two-body interaction between each pair of atoms. For each pair, the interaction potential was determined as a function of the distance between the two atoms, by performing an ab initio calculation using the MOLPRO code [17,23]. The present two-body potentials refer to the relaxed ground-state energy of the diatomic systems. No effect of the third atom on the two-body potentials was included. The trajectories of the three atomic cores (O, H, and Ar) were calculated using three pairs of two-body potentials (O-H, H-Ar, and O-Ar potentials). In this calculation, random initial conditions for the orientation and the position of the projectile were used. By repeating this calculation for a large number of collisions (~500 000), the energy and angular distributions of the ejected H fragments were determined. It was found that their energy distribution shows a double Gaussian structure similar to those found in the experiment. The narrow peak component is mainly due to collision events where the two-body character dominates, while the wide one is a manifestation of the effect of the third body (oxygen) on the kinematics of the H atom.

In a first attempt, no electronic excitation was introduced in the simulation. As a result, the simulated peak was found to be much narrower than the experimental one. For instance, the full width at half maximum (FWHM) of the simulated wider component did not exceed 40 eV, while it is experimentally found to be about 140 eV. Also, in contrast to the experimental results, no positive shift was found for the mean energy of the fragment ions emitted at small angles.

However, the formation of excited states may affect the angular and energy distributions of the H fragments. As mentioned above, a kinetic energy release may result from the decay of dissociative excited states and may increase the kinetic energy of the H fragment. Hence, to model the dissociative excitation of the $OH^+$ projectile, a further simulation was performed, in which a kinetic energy release was introduced. To do so, a velocity component along the OH axis, $\vec{v}_{KER}$, was added to the velocity of the H atom when the H-Ar distance is minimum. The corresponding KER ($\approx \frac{1}{2} m_H v_{KER}^2$) was assumed to be a random variable with a Gaussian distribution. Previous calculations of excitation energies for $OH^+$ [24] and measurements on the collision-induced Coulomb explosion of $H_2O$ molecules [25] show that the kinetic energy released in the breakage of an OH bond is on the order of 5 eV and can even exceed 20 eV when highly excited states are involved. Thus, as a reasonable compromise, a mean of 5 eV and a standard deviation of 4 eV were taken for the KER distribution in the present simulation (negative KER values were omitted).

As shown in Fig. 5, the introduced KER induces that the peaks are indeed broadened. A width of 100 eV is obtained for the wide component, which is not far from the experimental width (140 eV, with an instrumental resolution of 20 eV at the detection energy of 400 eV). Moreover, a positive energy shift (3 eV) is observed for this component at the smallest angles, but it is much smaller than the experimental one (10-20 eV). The qualitative agreement between the simulation and the experiment supports our interpretation in terms of a dissociative excitation of the molecular projectile which may contribute to the peak broadening and to the observed energy shift.

The cross section for H scattering as a function of the observation angle was also determined from the present three-body simulation, in which a KER is introduced. Except for a significant deviation at 3°, this simulation leads to a similar angular dependence compared to the calculation for elastic two-body scattering (Fig. 6). At emission angles larger or equal to 5°, deviations are less than 30% (Fig. 6). Likewise, above 5° the simulated cross section shows the same angular dependence as the experiment for $H^+$ emission (Fig. 6). The same feature applies for $H^-$ emission. These findings confirm that at scattering angles larger than a few degrees the presence of the third body (the oxygen atom) does not significantly affect the angular distribution of the scattered H. Moreover, these results support the conclusion that the distribution of the final charge states of the H fragments do not depend on the scattering angle, and thus, can be predicted by simple statistical laws.

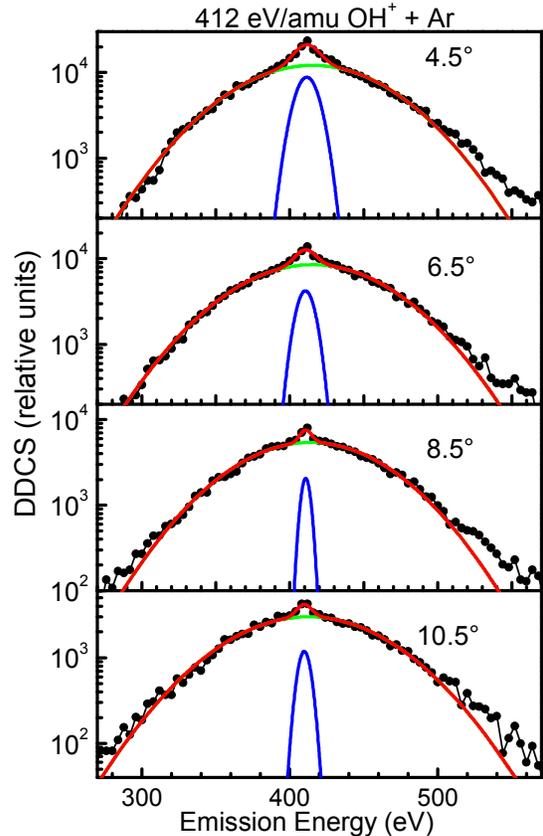

**Figure 5.** (Color online) Simulated energy distribution of the scattered H atoms (black dots) at different angles (see text). In the simulation, the impact parameter of the H center with respect to the Ar atom ranges from 0.05 to 2.25 a.u. The double Gaussian functions used for the fitting procedure are shown (blue curve: narrow component, green curve: wide component, red curve: sum).

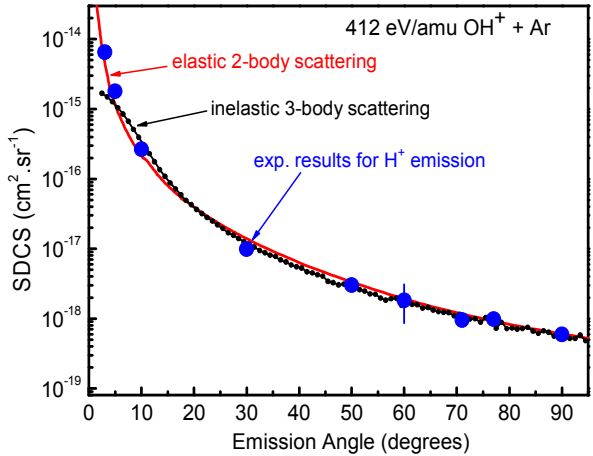

**Figure 6.** (Color online) Simulated angular distribution of the scattered H centers (black curve with symbols) in comparison with the experimental angular distribution of the emitted H$^+$ fragments (blue symbols). A normalization factor is applied to the simulated curve to match the experimental data, *i.e.* to convert the number of simulated events into cross section. Red curve: singly differential cross section for elastic two-body scattering of H on Ar (multiplied by 0.53, as in the lower part of Fig. 4).

**Conclusion**

We have shown that negative and positive hydrogen ions are emitted in OH$^+$ + Ar collisions with nearly the same angular dependence from very small to large scattering angles. For both H$^+$ and H$^-$ fragments, the measured emission cross sections are proportional to the calculated single differential cross section for elastic scattering of an incident H atom on an Ar atom. This feature is not only true for large emission angles (> 10°) due to violent binary collisions involving non-negligible momentum transfer at small impact parameter (< 1 a.u.), but also for small angles (< 10°) resulting from soft collisions at large impact parameters (> 1 a.u.). The comparison of the experimental data with the results of a three-body scattering simulation leads to a similar finding. Consistently, the ratio of H$^-$ to H$^+$ cross sections has been found to be constant in the whole investigated angular range. These findings provide evidence that the fractions of negative and positive H ions among all the emitted H fragments do not depend on the emission angle, and thus, barely depend on the momentum transferred to the proton during the collision. Hence, the earlier proposed statistical distribution of the final charge states [17] is confirmed.

The removal and emission of the H ion are driven by its scattering on the atomic target but the results cannot be explained without dissociative electronic excitation of the OH$^+$ projectile. Especially, the formation of an H fragment in a soft collision at large impact parameter requires a dissociative excited state, thus allowing the release of sufficient kinetic energy. A simulation in which excitation and dissociation of the projectile were modeled by introducing some KER reproduces qualitatively the experimental findings. This simulation suggests that the presence of the third body (the oxygen atom) may have a significant influence on the kinetic energy distribution of the emitted H fragment, while it barely affects its angular distribution. The KER due to dissociative excitation of the molecular projectile implies the addition of a velocity component in a random direction, which leads to a significant broadening of the energy distribution of the light H fragment. On the other hand, the effect of the KER becomes barely visible after integration over the emission energy, so that it is unobserved in the angular distribution.

For a deeper description of the electron capture and fragmentation processes, a more sophisticated *ab initio* calculation would be necessary. It is however likely that the transitions between intermediate quasi-molecular states cannot be followed due to the complexity of the system. But the present findings show that, due to the very limited number of their final states, the population of the outgoing fragments can be described by simple statistical laws.

**Acknowledgments**

The authors are grateful to Fabien Noury, Dr. Stéphane Guillous, and Dr. Patrick Rousseau

for their assistance during the measurements. This work was supported by the Transnational Access ITS-LEIF, granted by the European Project HPRI-CT-2005-026015, the Hungarian National Science Foundation OTKA (K73703 and K109440), the French-Hungarian Cooperation Program PHC Balaton (No. 27860ZL/TÉT_11-2-2012-0028), the French-Hungarian CNRS-MTA Cooperation number 26118 (ANION project) and the TÁMOP-4.2.2/B-10/1-2010-0024 project, cofinanced by the EU and the European Social Fund.